\documentclass{ragtime}
\usepackage{amsmath}
\usepackage{amssymb}
\usepackage{graphicx}
\usepackage{subfig}

\usepackage{adjustbox}
\usepackage{color}
\definecolor{darkgreen}{rgb}{0.0, 0.65, 0.31}
\usepackage[normalem]{ulem}

\title{Dynamical analysis approaches in spatially curved FRW spacetimes
}

\author[M. Kerachian, 
        G. Acquaviva  
        and G. Lukes-Gerakopoulos]
       {Morteza Kerachian\at{1,a} 
       Giovanni Acquaviva\at[]{1,b} 
         \splitauthors and Georgios Lukes-Gerakopoulos\at[]{2,c}\\
        \ins{1}Institute of Theoretical Physics, Faculty of Mathematics and Physics,\splitins[1]
        Charles University, CZ-180 00 Prague, Czech Republic\\
        \ins{2}Astronomical Institute of the Academy of Sciences of the Czech Republic,\splitins[1]
        Bo\v{c}n\'{i} II 1401/1a, CZ-141 00 Prague, Czech Republic\\
        \ins{a}\Email{kerachian.morteza@gmail.com} \\
        \ins{b}\Email{gioacqua@gmail.com}\\
        \ins{c}\Email{gglukes@gmail.com}}

\coentry{M. Kerachian, G. Acquaviva and G. Lukes-Gerakopoulos}%
       {Dynamical analysis approaches in spatially curved FRW spacetimes}

\graphicspath{{ker/}}

\begin{document}

\begin{abstract}
In this article, we summarize two agnostic approaches in the framework of spatially curved Friedmann-Robertson-Walker (FRW) cosmologies discussed in detail in \citep{Kerachian20,Kerachian19}. The first case concerns the dynamics of a fluid with an unspecified barotropic equation of state (EoS), for which the only assumption made is the non-negativity of the fluid’s energy density. The second case concerns the dynamics of a non-minimally coupled real scalar field with unspecified positive potential. For each of these models, we define a new set of dimensionless variables and a new evolution parameter. In the framework of these agnostic setups, we are able to identify several general features, like symmetries, invariant subsets and critical points, and provide their cosmological interpretation. 
\end{abstract}


\section{Introduction}

The dynamical system analysis is a powerful tool that has broad applications in different fields of science. Dynamics itself was introduced by Newton through his laws of motion and gravitation. These laws enabled Newton to tackle the two-body problem of the Earth's motion around the Sun. Later on, when scientists tried to address the three-body problem of the Earth, the Moon and the Sun, they found it was too complicated to tackle it quantitatively. In the late 19th century, Henry Poincar\'{e} suggested that celestial mechanics could be studied by considering qualitative features of a system rather than quantitative founding in this way the branch of dynamical systems~\citep{strogatz2018nonlinear}. In the context of cosmology dynamical systems analysis
allows us to view the global evolution of a model, from its start near the initial singularity to its late-time evolution~\citep{wainwright2005cosmological}.

The observations indicate that the universe is homogeneous and isotropic \citep{aghanim2018planck}, which makes the Friedmann-Robertson-Walker (FRW) spacetime the relevant metric to model its evolution. Even if the universe appears to be spatially flat, considering a non-zero spatial curvature is still observationally viable and might help in alleviating some cosmological tensions \citep{ryan2019baryon,di2019planck}. Therefore, in our work we used spatially curved FRW metrics. 

According to \cite{refId0}, the total energy density of the universe consist of $\sim68.5\%$ dark energy, $\sim26.5\%$ cold dark matter, and $\sim5\%$ baryonic matter. There are three main approaches in order to understand the physics behind the dominant substance of the universe, i.e. the dark energy: the constant vacuum energy or cosmological constant, non-constant vacuum energy or scalar fields, and modified gravities. The cosmological constant scenario, expressed by the $\Lambda$CDM model, is considered as the standard model for describing dark energy, but since it suffers from several issues~\citep{carroll2001cosmological,bahamonde2018dynamical} there are plenty of models that compete with it. In this work, we explore the dynamics of two such models in a rather general framework.

The first type of models we analyse concerns the dynamics of barotropic fluids with $\epsilon \ge 0$ in spatially curved FRW without specifying the EoS \citep{Kerachian20}. We allow the pressure $P$ of the fluid to attain negative values in order to be able to describe cosmological models with accelerated expansion. In these models the speed of sound of the fluid is not necessarily less than the speed of light, which implies exotic EoS. 

The second type of models we analyse concerns a curved FRW geometry non-minimally coupled to a scalar field with generic positive potential \citep{Kerachian19}. A similar analysis has been performed by \cite{Hrycyna2010} in the presence of matter for flat FRW. Our formulation allows for several improvements in the aforementioned analysis by considering a generic spatially curved FRW model and a more general scalar field potential.

\section{The dynamical system for Barotropic fluids}
\label{sec:sys}

The Friedmann and the Raychaudhuri equations for a FRW cosmology with only one fluid component are given by
\begin{equation}
 H^2 + \frac{k}{a^2} =\frac{\epsilon}{3}\, , \qquad
 2\,\dot{H} +3\, H^2 + \frac{k}{a^2}= -P\, ,\label{eq:raych}
\end{equation}
 respectively and the continuity equation for the energy density reads
\begin{equation}\label{eq:state}
\dot{\epsilon} +3\, H (P+\epsilon)=0\,. 
\end{equation}
In these equations, $\epsilon$ is the energy density, $P$ is the pressure of the barotropic fluid, $k$ is the spatial curvature, $a$ is the scale factor, $H=\frac{\dot{a}}{a}$ is the Hubble expansion rate and $\dot{~}$ denotes derivative with respect to the coordinate time.

By introducing the normalization $D^2=\displaystyle H^2+|k|/a^2$, we are able to present well-defined dimensionless variables, i.e. the variables which are valid for $k>0$ and $\displaystyle k\leq 0$. These new dimensionless variables are
\begin{equation}
 \Omega_{\epsilon} = \frac{\epsilon}{3\,D^2},\quad  \Omega_H = \frac{H}{D}, \quad \Omega_{P} =\frac{P}{D^2},\quad \Omega_{\partial P} = \frac{\partial P}{\partial \epsilon}, \quad \Gamma = \frac{\partial^2P}{\partial\epsilon^2}\epsilon.
 \label{eq:secvar2}
\end{equation}

In order to investigate the evolution of the dimensionless variables. we define a new evolution parameter $\tau$  as $d \tau = D dt$. This new evolution parameter is well-defined during the whole cosmic evolution. Taking the derivative of the dimensionless variables with respect to $\tau$ provides the autonomous system
\begin{align}
 \Omega_{\epsilon}' &= -\Omega_H \left[\Omega_p+\Omega_{\epsilon} \left(3+2\left(\frac{\dot{H}}{D^2} + \Omega_H^2 - 1\right) \right) \right]\label{eq:omegaep} \, ,\\
 \Omega_H' &= \left( 1 - \Omega_H^2\right)\, \left( \frac{\dot{H}}{D^2} + \Omega_H^2\right) \, , \label{eq:omegahp}\\
 \Omega_P' &= -\Omega_H \left[ 3 \Omega_{\partial P}\left( \Omega_P+3 \Omega_{\epsilon} \right)  +2 \Omega_P\left( \frac{\dot{H}}{D^2} + \Omega_H^2 - 1\right)\right] \, ,
 \label{eq:omegap}\\
  \Omega_{\partial P}' & =-\Omega_H\left(\frac{\Omega_P}{\Omega_{\epsilon}}+3\right)\,\Gamma\label{eq:Csp} \, .
\end{align}

\paragraph{\textbf{Positive curvature:}} For positive curvature $k>0$, in terms of the new variables the Friedmann and Raychaudhuri equations~\eqref{eq:raych} become respectively
\begin{equation} \label{eq:DeRaychPos}
\Omega_{\epsilon}=1, \qquad \frac{\dot{H}}{D^2}=-\frac{1}{2}\left(\Omega_P+1\right)-\Omega_H^2 \, .
\end{equation}

\paragraph{\textbf{Non-positive curvature:}}\label{sec:nprych}
For the non-positive spatial curvature $k\leq0$, in terms of the new variables the Friedmann and Raychaudhuri equations~\eqref{eq:raych} become respectively
\begin{equation}\label{eq:DeRaychNPos}
\Omega_{\epsilon}= 2\, \Omega_H^2-1,\qquad \frac{\dot{H}}{D^2}=-\frac{1}{2}\left(\Omega_P+1\right)+\left(1-2\Omega_H^2\right).
\end{equation}
From the definition of $\Omega_H$ we have $\Omega_H^2\le 1$ and from the assumption $\epsilon \ge 0$, we get that $0\le \Omega_\epsilon \le 1$ and $\frac{1}{2}\leq\Omega_H^2\leq1$.

\subsection{Critical points and their interpretation}\label{sec:CritPoi_ss}

The next step is to investigate the critical points ( i.e. those points for which $\mathbf{\Omega}'=0$) of the autonomous system~\eqref{eq:omegaep}-~\eqref{eq:Csp} and their stabilities. Once the critical points are determined, we can look for their cosmological interpretation. To do that a useful tool is the {\it deceleration parameter}
\begin{align}
 q =-1-\frac{\dot{H}}{H^2}=-1-\Omega_H^{-2}\, \frac{\dot{H}}{D^2}\, ,
\end{align}
in which we used the definition of $\Omega_H$.  

\paragraph{\textbf{Two de Sitter critical lines:}} There are two critical lines with a de Sitter behavior located at $\{\Omega_{\epsilon},\Omega_{H},\Omega_{P},\Omega_{\partial P}\}=\{1,\pm 1,-3, \forall \}$. The critical line with $\Omega_H=1$ (called $A_+$) has the typical cosmological constant behaviour $(q=-1)$ and its eigenvalues are
\begin{equation} \label{eq:eigap}
\{\lambda^{A_{+}}_{i} \}=\{-2 , 0 ,-3\left(1+\Omega_{\partial P}\right)  \},
\end{equation}
while the critical line with $\Omega_{H}=-1$ (called $A_{-}$) describes an exponentially shrinking universe $(q=-1)$ and its eigenvalues are
\begin{equation}\label{eq:eigam}
\{\lambda^{A_{-}}_{i} \}=\{2 , 0 ,3\left(1+\Omega_{\partial P}\right)  \}.
\end{equation}

Eq.~\eqref{eq:eigap} and  Eq.~\eqref{eq:eigam} imply that for $\Omega_{\partial P} < -1$ the critical points along the lines $A_\pm$ are saddle points. However, for $\Omega_{\partial P} \geq -1$ the stability of the points along $A_\pm$ can not be determined even by the center manifold theorem. To discuss their stability numerical examples for specific $\Gamma$ have to be employed.  

\paragraph{\textbf{Static universe critical line:}}\label{sec:CPB2} For positive spatial curvature, there is a critical line (called $B$) located at $\{\Omega_{\epsilon},\Omega_{H},\Omega_{P},\Omega_{\partial P}\}=\{1,0,-1,\forall \}$. This critical line describes a static universe, i.e $a = \textrm{const.}$ and its eigenvalues are
\begin{equation}\label{eq:eigb}
\{\lambda^{B}_{i} \}=\{0, -\sqrt{1+3 \Omega_{\partial P}} ,\sqrt{1+3 \Omega_{\partial P}} \}.
\end{equation}
Eq.~\eqref{eq:eigb} implies that for $1+3 \Omega_{\partial P} > 0$, the critical points along the line $B$ are saddle; for $1+3 \Omega_{\partial P} < 0$ these points are center; for $\Omega_{\partial P}=-1/3$ the corresponding points are degenerate and all eigenvalues are zero. Since the center manifold theory cannot be employed, we rely on a numerical inspection which shows that this point is marginally unstable.

For negative curvature, there is another critical line (called $\Bar{B}$) corresponding to a static universe  located at $\{\Omega_{\epsilon},\Omega_{H},\Omega_{P},\Omega_{\partial P}\}=\{-1,0,1,\forall \}$,  but as discussed in Sec.~\ref{sec:nprych}, $\Omega_\epsilon<0$ cases are not part of our study.

\subsection{General features of $\Gamma$: invariant subsets and critical points}
\label{sec:roottrack}

In this section let us assume that the function $\Gamma$ has roots $\displaystyle \Tilde{\Omega}_{\partial P}$: this allows invariant subsets lying on $\{\Omega_{H},\Omega_{P}\}$ planes. For each root of $\Gamma$, we get a pair of critical points $C_\pm$ located at $\displaystyle \{\Omega_{H},\Omega_{P}\}=\{\pm 1,3\, \Tilde{\Omega}_{\partial P}\}$. Note that, for any new invariant subset $\{\Omega_{H},\Omega_{P}\}$ there might be an intersection with the critical lines $A_\pm$ and $B$; for simplicity we denote these resulting critical points  with the same name as the respective critical lines.

The scale factor for the critical point $C_+$ grows as $\displaystyle a \sim (t-t_0)^{\frac{2}{3\,(\Tilde{\Omega}_{\partial P}+1)}} $, while for the critical point $C_-$ it decreases as $\displaystyle a \sim (t_0-t)^{\frac{2}{3\,(\Tilde{\Omega}_{\partial P}+1)}} $. At these points the deceleration parameter reduces to $  q=\frac{1}{2}(3\,\Tilde{\Omega}_{\partial P}+1)$. $C_\pm$ according to $q$ represent an accelerated universe when $ \Tilde{\Omega}_{\partial P}< -\frac{1}{3}$ and a decelerated one when $ \Tilde{\Omega}_{\partial P}> -\frac{1}{3}$.

\begin{figure}[htp]
\begin{center}
\captionsetup[subfloat]{position=top} 
 {\subfloat[$\Tilde{\Omega}_{\partial P}=0.5$]{\includegraphics[width=0.31\textwidth]{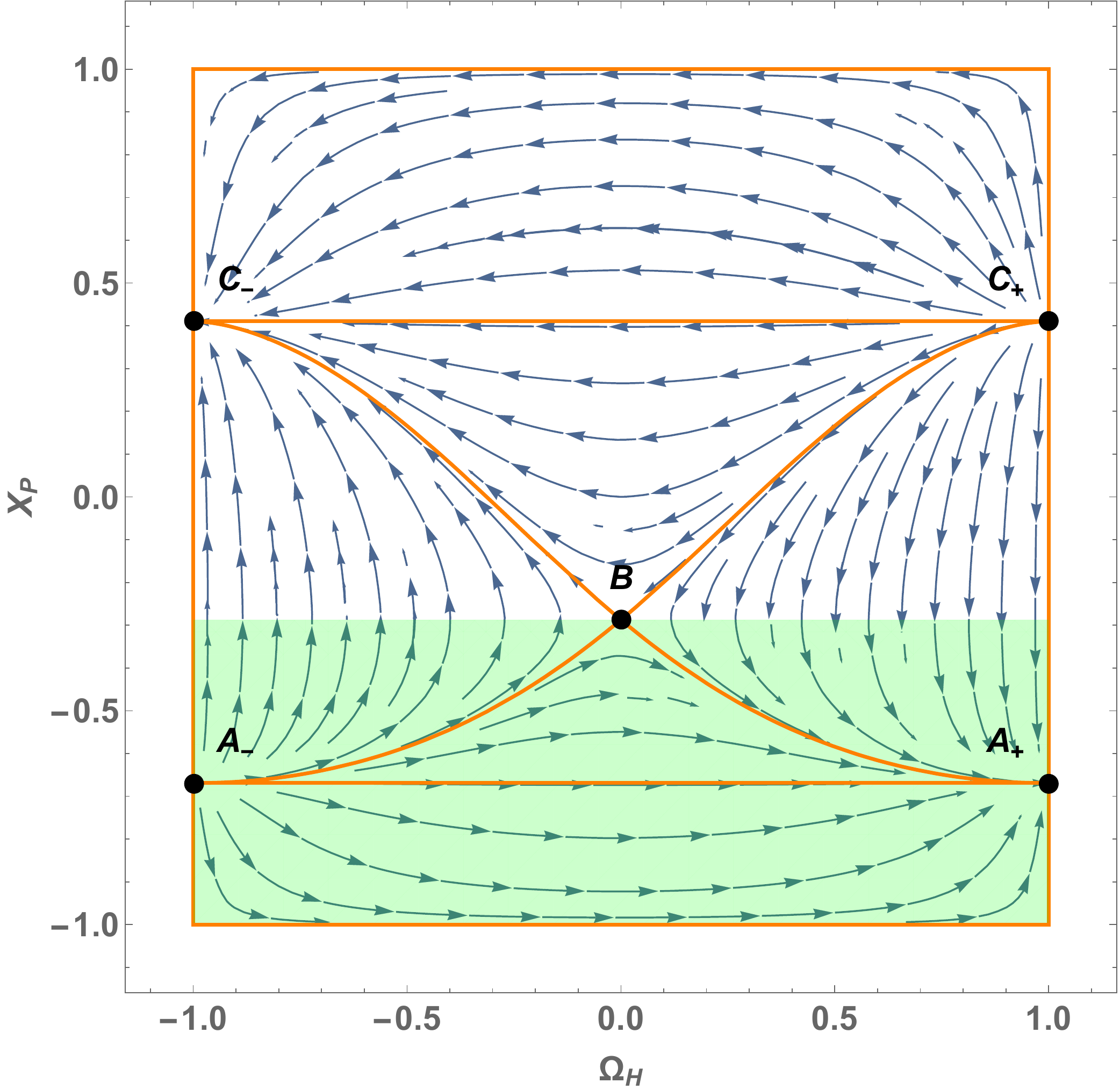}\label{fig:gen1}}}
{  \subfloat[$\Tilde{\Omega}_{\partial P}=-0.6$]{\includegraphics[width=0.31\textwidth]{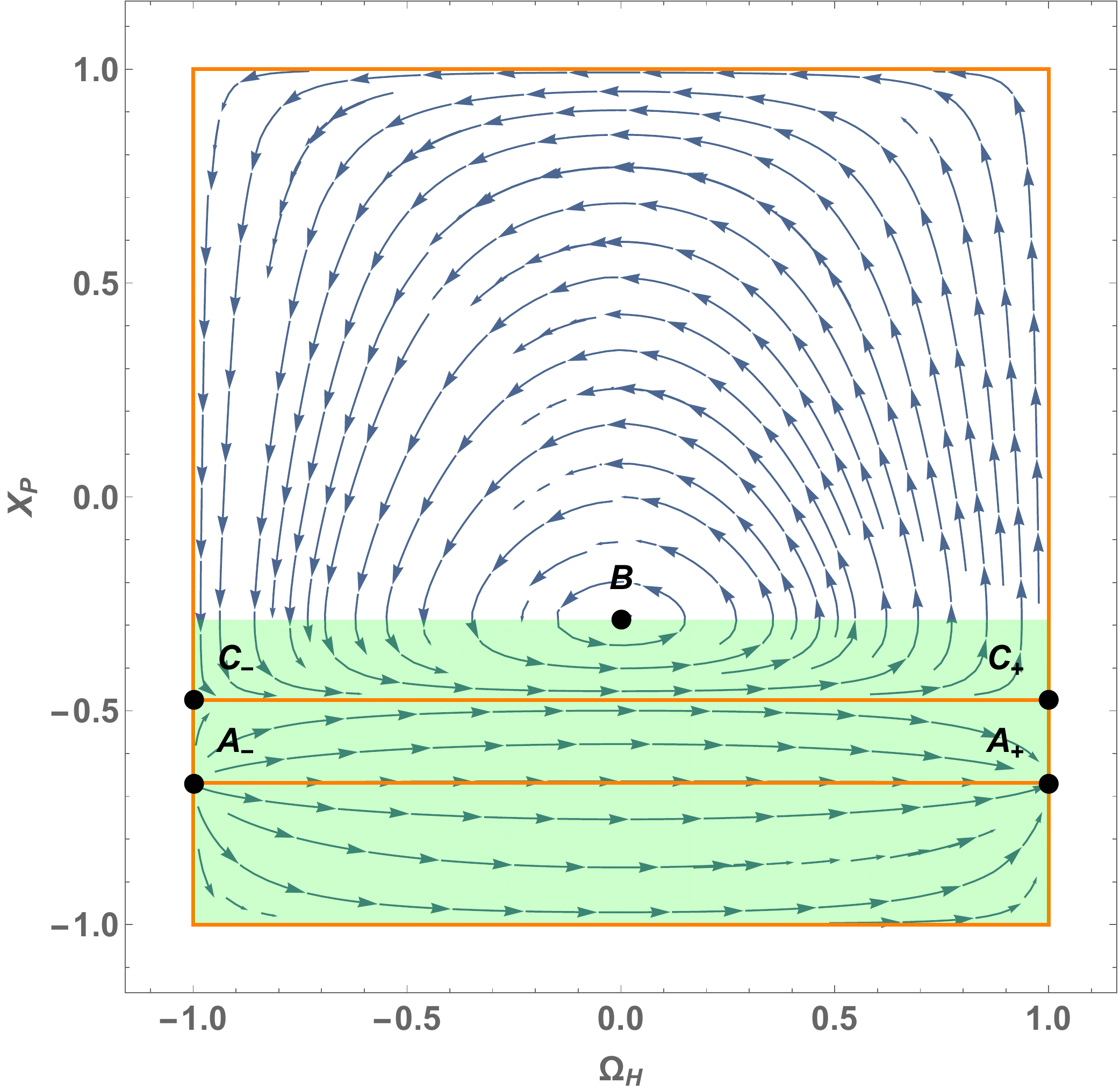}\label{fig:gen2}}}
{  \subfloat[$\Tilde{\Omega}_{\partial P}=-1.4$]{\includegraphics[width=0.31\textwidth]{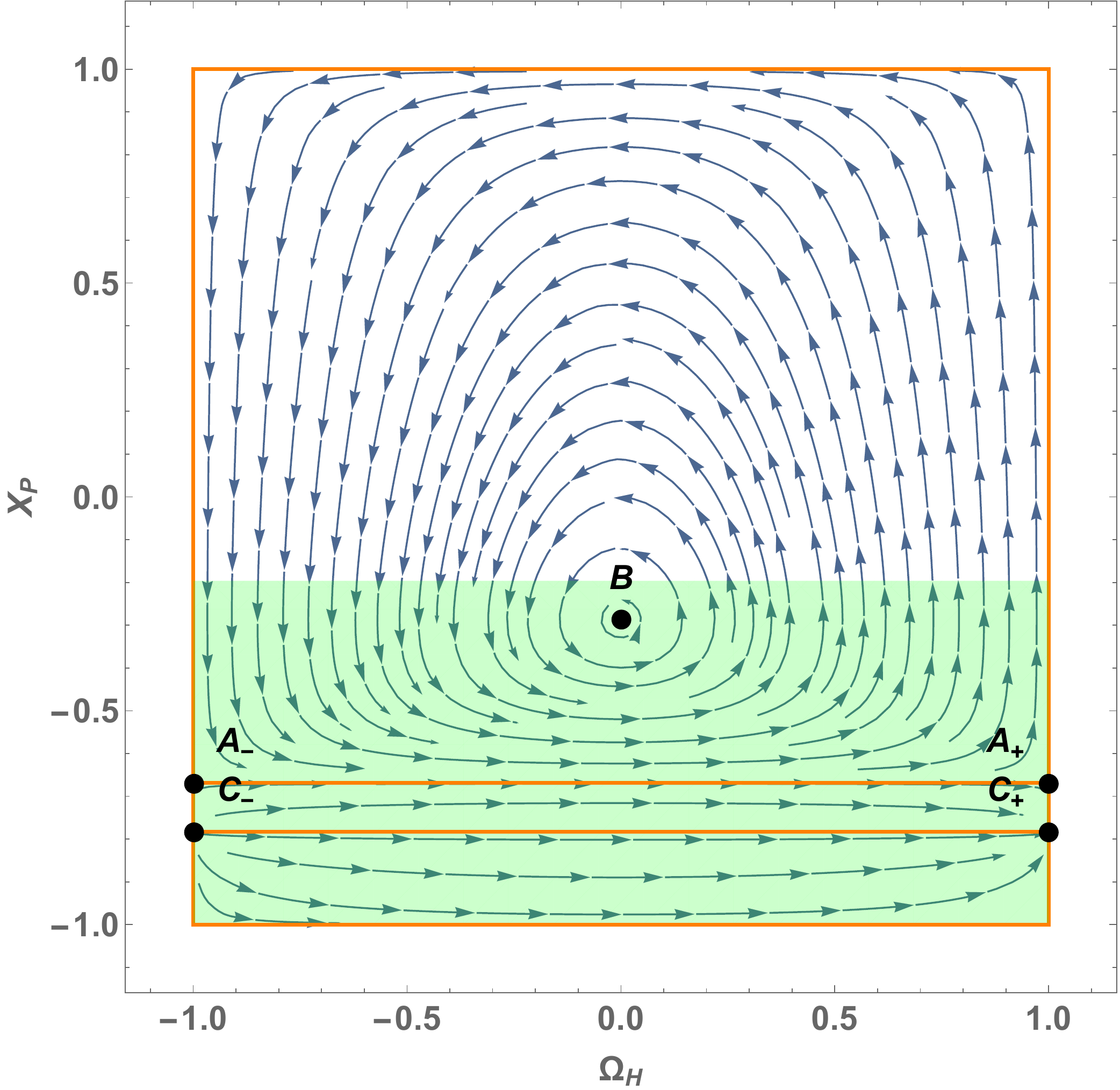}\label{fig:gen3}}}   
\end{center}
    \caption{Invariant subsets for positive spatial curvature and $\zeta=0.3$ plotted for three representative values of $\Tilde{\Omega}_{\partial P}$ in the ranges given in Sec.~\ref{sec:rootpos}. The orange thick lines are the separatrices of the system and the green shaded regions denote the part of the variable space where the universe is accelerating.}\label{fig:genp}
\end{figure}

\begin{figure}[htp]
    \centering
    \captionsetup[subfloat]{position=top} 
{\subfloat[$\Tilde{\Omega}_{\partial P}=0.5$]{\includegraphics[width=0.31\textwidth]{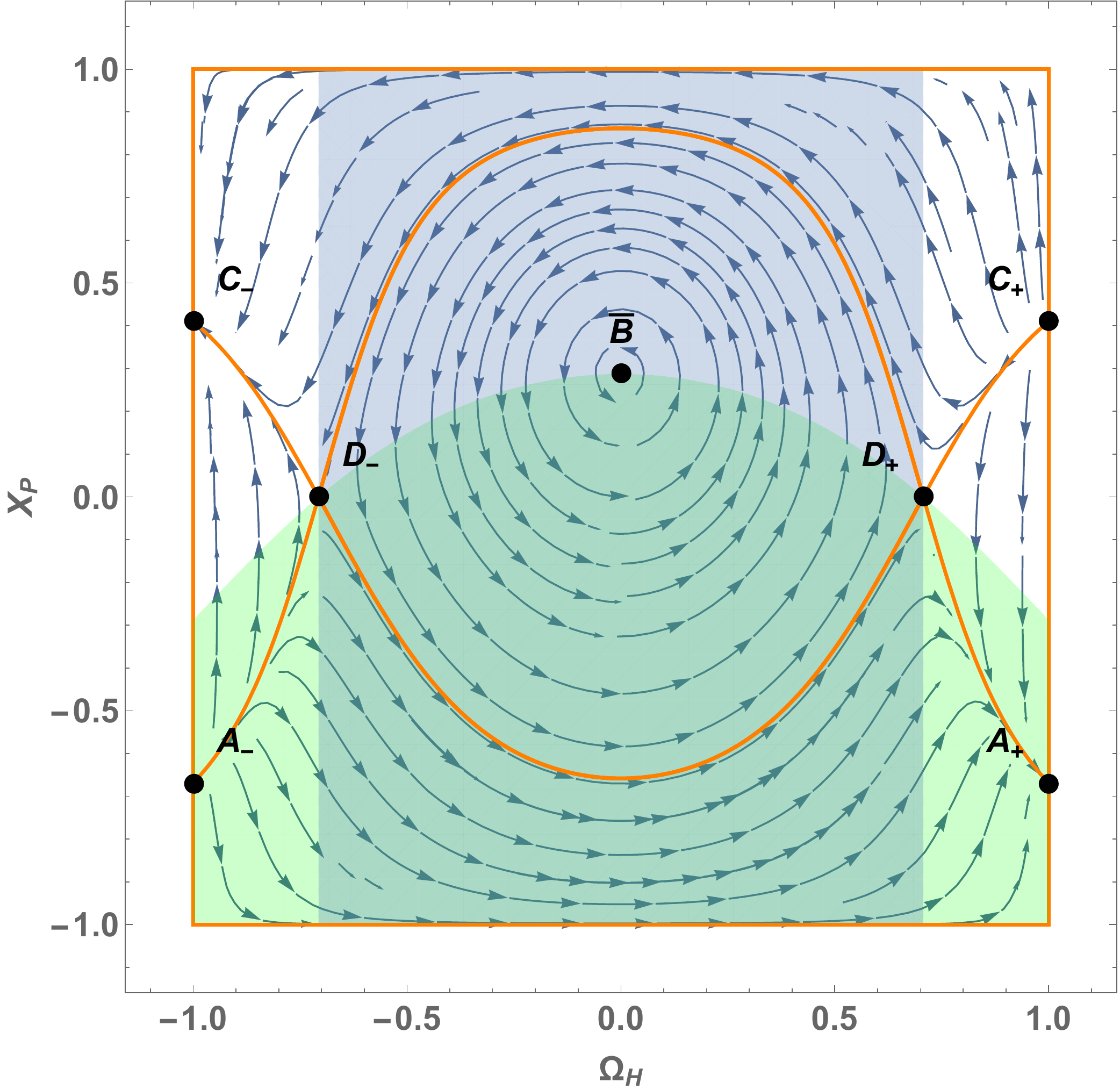}\label{fig:genn1}}}
{  \subfloat[$\Tilde{\Omega}_{\partial P}=-0.6$]{\includegraphics[width=0.31\textwidth]{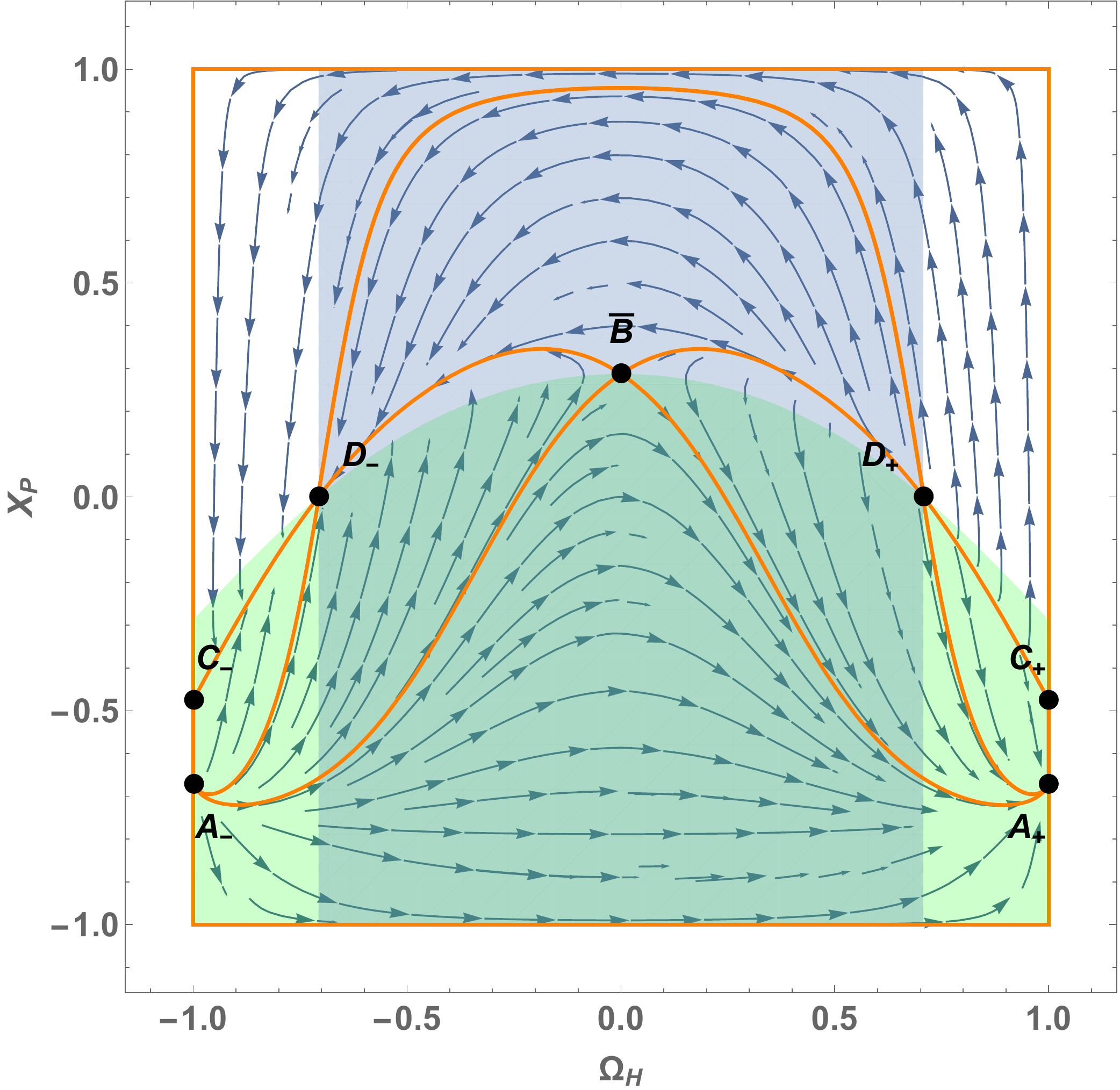}\label{fig:genn2}}}
{  \subfloat[$\Tilde{\Omega}_{\partial P}=-1.4$]{\includegraphics[width=0.31\textwidth]{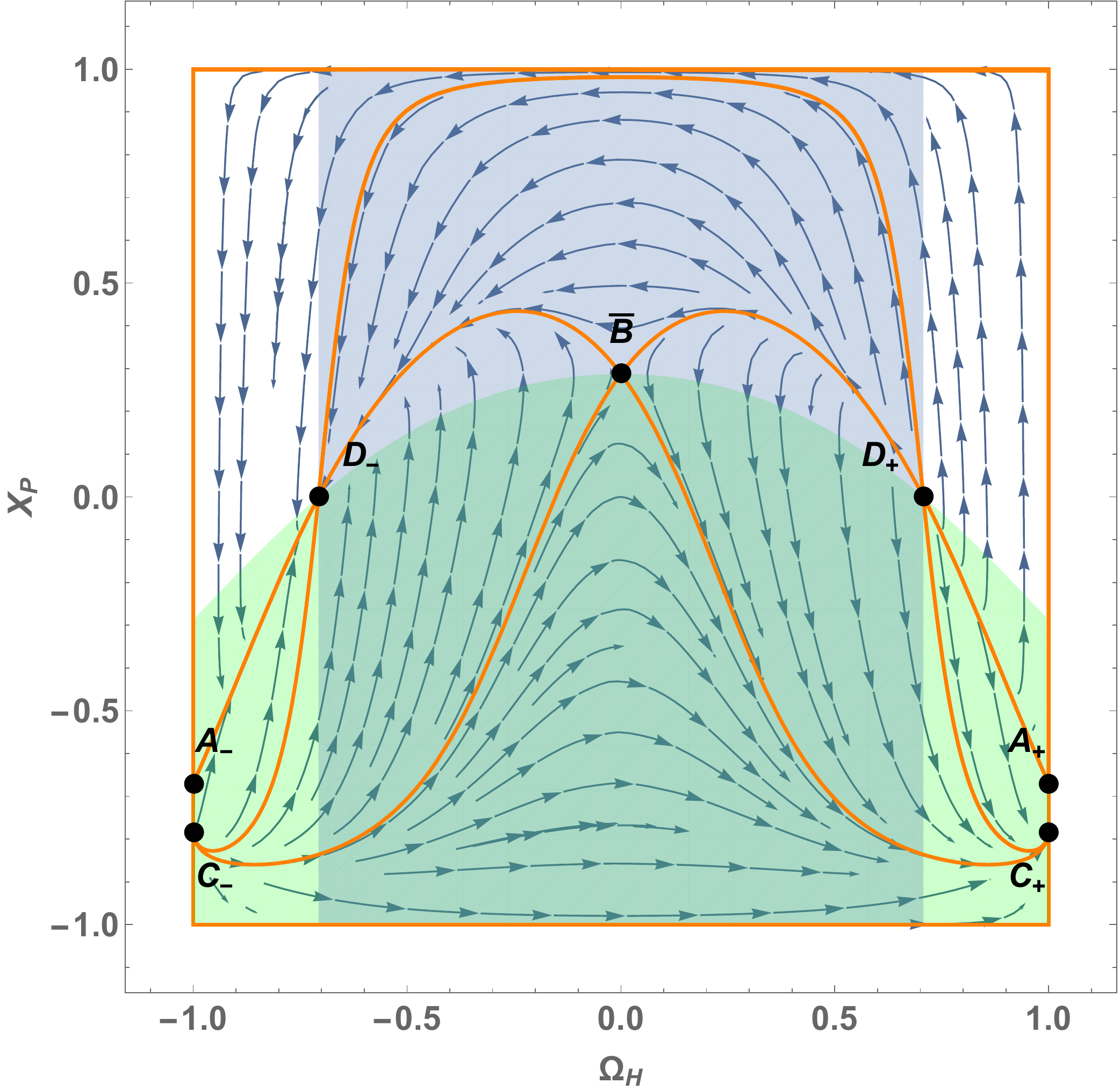}\label{fig:genn3}}}
    \caption{Invariant subsets for negative spatial curvature and $\zeta=0.3$ plotted for three representative values of $\Tilde{\Omega}_{\partial P}$ in the ranges given in Sec.~\ref{sec:nproots}. The orange thick lines are the separatrices. The  blue shaded areas are the regions excluded by our assumption that $\Omega_\epsilon>0$. The green shaded region are the part of the variable space where we have accelerating universe.
    }\label{fig:genn}
\end{figure}

The points $C_\pm$ have eigenvalues 
\begin{equation}
    \{\lambda^{C_\pm}_{i} \}=\{\pm3 \,(1+\Tilde{\Omega}_{\partial P}),\pm(1+3\,\Tilde{\Omega}_{\partial P}) \}.\label{eq:eigcp}
\end{equation}
Based on these eigenvalues on the invariant subset $\{\Omega_H,\Omega_P\}$ and one can see that for $-\frac{1}{3}<\Tilde{\Omega}_{\partial P}$ point $C_+$ ($C_-$) is a source (sink). For the case $-1< \Tilde{\Omega}_{\partial P}<- \frac{1}{3}$ instead $C_\pm$ are saddle. Finally, for $\Tilde{\Omega}_{\partial P}<-1$ point $C_+$ ($C_-$) is a sink (source). These points can be seen in the examples shown in Figs.~\ref{fig:genp} and~\ref{fig:genn}.

Since the stability of the critical points ($A_\pm$, $B$,  and $C_\pm$) of the system depends on the value of $\Tilde{\Omega}_{\partial P}$, we split our analysis into the following three ranges
\begin{align}
    -\frac{1}{3}<\Tilde{\Omega}_{\partial P}, \qquad -1< \Tilde{\Omega}_{\partial P}<- \frac{1}{3}, \qquad \Tilde{\Omega}_{\partial P}<-1.
\end{align}
and we are going to depict the invariant subset $\{\Omega_H, \Omega_P\}$ in these ranges. In Figs.~\ref{fig:genp},~\ref{fig:genn} we choose one representative value of $\Tilde{\Omega}_{\partial P}$ for each range, since the topology of the trajectories is independent of the specific value inside each range. For simplicity we assume that the function $\Gamma$ has only one root.

In order to be able to investigate the asymptotic behaviour of $\Omega_P$, i.e. $\Omega_P=\pm \infty$, in Figs.~\ref{fig:genp} and~\ref{fig:genn} we used the transformation 
\begin{equation}\label{eq:xp}
    X_P=\frac{\zeta\Omega_P}{\sqrt{1+\zeta^2\Omega_P^2}} \in [-1,1],
\end{equation}
where $\zeta>0$ is just a constant rescaling parameter. The evolution equation for this variable for positive curvature becomes 
\begin{align}\label{eq:xpp}
    X_P'= \frac{\Omega_H}{\zeta} \sqrt{1-X_P^2}
   \left( X_P+3\,\zeta\,\sqrt{1-X_P^2}\right)\,\left(X_P-3\,\zeta\,\Omega_{\partial P}\,\sqrt{1-X_P^2}\right),
\end{align}
while for the non-positive curvature becomes 
\begin{align}\label{eq:xppn}
      X_P' &= \frac{\Omega_H}{\zeta} \sqrt{1-X_P^2}\,( 9\,\zeta^2\,\Omega_{\partial P}\,(1-2\,\Omega_H^2)\,(1-X_P^2)+\nonumber\\
  &  \zeta\, X_P\,\sqrt{1-X_P^2}\,(1-3\,\Omega_{\partial P}+2\,\Omega_H^2)+X_P^2)),
\end{align}
which along with the Eq.~\eqref{eq:omegahp} define the compactified systems.

\paragraph{\textbf{Positive curvature:}} \label{sec:rootpos}
Fig.~\ref{fig:genp} shows the invariant subsets $\lbrace \Omega_H, X_P \rbrace$ for the positive curvature, on which two additional invariant subsets are located at $\Omega_P=-3$ and $\Omega_P=3\,\Tilde{\Omega}_{\partial P}$. 

\paragraph{\textbf{Non-positive curvature}}\label{sec:nproots} 
For the non-positive curvature there are additional critical points once we consider the roots $\Gamma( \Tilde{\Omega}_{\partial P})=0$. The locations of these critical points are $\{\Omega_{H},\Omega_{P}\}=\{\pm \frac{1}{\sqrt{2}},0\}$ and they represent a Milne universe, since the deceleration parameter $q=0$ and the scale factor evolves as $a=\pm \mid k \mid (t+c_1)$ for $\Omega_H=\pm\,\frac{1}{\sqrt{2}}$.

The critical point with $\Omega_H=\frac{1}{\sqrt{2}}$ denoted as $D_{+}$ has eigenvalues 
\begin{equation}
    \{\lambda^{D_+}_{i} \}=\{\sqrt{2}, -\frac{\sqrt{2}}{2}\,\left(1+3\,\Tilde{\Omega}_{\partial P}\right) \},\label{eq:eigdp}
\end{equation}
in the invariant subset $\lbrace \Omega_H,\Omega_P\rbrace$, whiles the critical point denoted as $D_-$ has eigenvalues
\begin{equation}
    \{\lambda^{D_-}_{i} \}=\{-\sqrt{2}, \frac{\sqrt{2}}{2}\,\left(1+3\,\Tilde{\Omega}_{\partial P}\right) \}.\label{eq:eigdm}
\end{equation}
Eqs.~\eqref{eq:eigdp} and~\eqref{eq:eigdm} show that for $-\frac{1}{3}<\Tilde{\Omega}_{\partial P}$ the critical points $D_\pm$ are saddles, while for $-\frac{1}{3}>\Tilde{\Omega}_{\partial P}$, $D_+$ is a source and $D_-$ is a sink. 

\section{The dynamical system for Non-minimally coupled scalar field}

The action of a scalar field non-minimally coupled to gravity reads
\begin{equation}\label{eq:actionnon-minimal}
    S= \int d^4x \sqrt{-g} \left( \frac{R}{2}+ \mathcal{L}_\psi \right),
\end{equation}
where $\mathcal{L}_\psi$ is the Lagrangian for the scalar field $\psi$:
\begin{equation}
    \mathcal{L}_\psi= - \frac{1}{2}\,\left( g^{\mu \nu}\, \partial_\mu \psi\, \partial_\nu \psi + \xi R \psi^2 \right)- V(\psi),
\end{equation}
and $V(\psi)$ is a scalar field potential. 

By variation of the action~\eqref{eq:actionnon-minimal} with respect to $g_{\mu \nu}$, we arrive to the Einstein field equations
\begin{equation}\label{eq:EFEnon-minimal}
    R_{\mu \nu}- \frac{1}{2} R \,g_{\mu \nu}= T_{\mu \nu}^\psi.
\end{equation}
where  the stress-energy tensor $T_{\mu \nu}^\psi$ for the non-minimally coupled scalar field reads 
\begin{align}
    T_{\mu \nu}^\psi= (1-2 \,\xi) \nabla_\mu \psi \,\nabla_\nu \psi +\left(2\,\xi-\frac{1}{2} \right) g_{\mu \nu} \nabla^\alpha \psi\,\nabla_\alpha \psi- V(\psi) \,g_{\mu \nu}\nonumber\\
    +\xi \left( R_{\mu \nu}- \frac{1}{2} g_{\mu \nu} R\right)\, \psi^2+ 2\,\xi \psi \left(g_{\mu \nu} \,\nabla^\alpha\,\nabla_\alpha- \nabla_\mu\,\nabla_\nu \right) \psi.
\end{align}
By variation of the action with respect to the scalar field~$\psi$ we get the Klein-Gordon equation 
\begin{equation}\label{eq:non-minkleingord}
    \nabla _\mu \, \nabla ^\mu \, \psi -\xi R \psi- \frac{\partial V(\psi)}{\partial \psi}=0.
\end{equation}

The Friedmann and the Raychaudhuri equations for the non-minimally coupled scalar field in the FRW background read
\begin{align} 
 3\, \left( H^2 + \frac{k}{a^2} \right)= \epsilon_\psi ,\quad
    \left( 2\,\dot{H} +3\, H^2 + \frac{k}{a^2} \right) = - P_\psi,
    \end{align}\label{eq:non-minimalfried}
respectively, while the Klein-Gordon equation reads
\begin{equation}\label{eq:non-minimalklein}
    \ddot{\psi} + 3\, H\, \dot{\psi} + \partial_{\psi} V + 6\, \xi\, \psi\,  \left(\dot{H} + 2\, H^2 + \frac{k}{a^2} \right) = 0.
\end{equation} 
Here the $\epsilon_\psi$ and $P_\psi$ are defined as
\begin{align}
    \epsilon_\psi &= \frac{1}{2}\, \dot{\psi}^2 + V(\psi)+3\,\xi\,\psi\,\left(2\, H\, \dot{\psi} +\psi \left( H^2 + \frac{k}{a^2} \right)\right),\label{eq:non-minepsi}\\
    P_\psi&= \left(1-4\, \xi\right)\frac{1}{2}\, \dot{\psi}^2 - V(\psi)- \xi \left( 4\, H\, \psi\, \dot{\psi}+ 2\,\psi\, \ddot{\psi}+ \psi^2\,\left( 2\,\dot{H} +3\, H^2 + \frac{k}{a^2} \right)\right).\label{eq:non-minP}
\end{align}

We define a set of dimensionless variables which are well-defined for positive and non-positive curvatures:
\begin{align}
 \Omega = \frac{\psi}{\sqrt{1+\xi\, \psi^2}},\quad\Omega_H = \frac{H}{D},\quad\Omega_{\psi} = \frac{\dot{\psi}}{\sqrt{6}\, D},\\
 \Omega_V = \frac{\sqrt{V}}{\sqrt{3}\, D},\quad\Omega_{\partial V} = \frac{\partial_\psi V}{V},\quad\Gamma = \frac{V \cdot\partial^2_\psi V}{(\partial_\psi V)^2}
 \label{eq:var2}
\end{align}
where $D^2=H^2+\frac{|k|}{a^2}$. Similarly as for the dynamical system in Sec.~\ref{sec:sys}, for these dimensionless variables the evolution parameter $\tau$  is defined as $d \tau = D dt$.  By taking derivatives of the dimensionless variables with respect to the evolution parameter we get
\begin{align}
 \Omega' &= \sqrt{6}\ \Omega_\psi\ \left( 1-\xi\, \Omega^2 \right)^{3/2} \label{eq:omega}\\
 \Omega_H' &= \left( 1 - \Omega_H^2 \right)\, \left( \frac{\dot{H}}{D^2} + \Omega_H^2 \right) \label{eq:omegah}\\
 \Omega_\psi' &= \frac{\ddot{\psi}}{\sqrt{6}\, D^2} - \Omega_\psi\, \Omega_H\, \left( \frac{\dot{H}}{D^2} + \Omega_H^2 - 1 \right) \label{eq:omegapsi}\\
 \Omega_V' &= \Omega_V\, \left[ \sqrt{\frac{3}{2}}\ \Omega_{\partial V}\ \Omega_{\psi} - \Omega_H\, \left( \frac{\dot{H}}{D^2} + \Omega_H^2 - 1 \right) \right] \label{eq:omegav}\\
 \Omega_{\partial V}' &= \sqrt{6}\ \Omega_{\partial V}^2\ \Omega_\psi\, \left( \Gamma - 1 \right)\, , \label{eq:omegadv}
\end{align}
where $\Gamma=V\, \cdot\, \partial^2_{\psi} V / \left( \partial_{\psi} V \right)^2$ which is the so-called tracker parameter. This autonomous system of equations differs only in the $\frac{\ddot{\psi}}{\sqrt{6}\ D^2}$ and $\frac{\dot{H}}{D^2}$ terms for $k>0$ and $k\leq0$. Namely for positive curvature we get from Klein-Gordon and Raychaudhuri equations 
\begin{align}
&\frac{\ddot{\psi}}{\sqrt{6}\ D^2} = - 3\ \Omega_H\ \Omega_\psi - \sqrt{\frac{3}{2}}\ \Omega_{\partial V}\ \Omega_V^2 \nonumber - \frac{\sqrt{6}\ \xi\ \Omega}{\sqrt{1-\xi\ \Omega^2}} \left(  \frac{\dot{H}}{D^2} + \Omega_H^2 + 1 \right)\, ,\\
& \frac{\dot{H}}{D^2} + \Omega_H^2 + 1 = - \frac{1}{1-2\ \xi\ (1-3\, \xi)\ \Omega^2}\ \Bigg\{-\frac{1}{2}\left( 1-2\, \xi\, \Omega^2 \right) \nonumber\\ 
 &+ \xi\, \Omega\, \sqrt{1-\xi\, \Omega^2} \left( \sqrt{6}\, \Omega_H\, \Omega_\psi+3\, \Omega_{\partial V}\, \Omega_V^2 \right)\nonumber + \frac{3}{2}\left(1-\xi\, \Omega^2\right) \Big[ (1-4\, \xi)\, \Omega_\psi^2-\Omega_V^2 \Big]  \Bigg\}\, , \nonumber
\end{align}
while for non-positive curvature these equations read
\begin{align*}
&\frac{\ddot{\psi}}{\sqrt{6}\ D^2} = - 3\ \Omega_H\ \Omega_\psi - \sqrt{\frac{3}{2}}\ \Omega_{\partial V}\ \Omega_V^2 + \frac{\sqrt{6}\ \xi\ \Omega}{\sqrt{1-\xi\ \Omega^2}} \left(  1- \frac{\dot{H}}{D^2} - 3\, \Omega_H^2 \right)\,  ,\\
& \frac{\dot{H}}{D^2} +\Omega_H^2 = \frac{1}{2} - \Omega_H^2 \nonumber + \frac{1}{1-2\ \xi\ (1-3\, \xi)\ \Omega^2}\ \Bigg\{ 3\, \xi^2\, \Omega^2\, \left( 1-2\, \Omega_H^2 \right)\nonumber \\ 
 &- \xi\, \Omega\, \sqrt{1-\xi\, \Omega^2} \left( \sqrt{6}\, \Omega_H\, \Omega_\psi+3\, \Omega_{\partial V}\, \Omega_V^2 \right)- \frac{3}{2}\left(1-\xi\, \Omega^2\right) \Big[ (1-4\, \xi)\, \Omega_\psi^2-\Omega_V^2 \Big] \Bigg\}\, .
\end{align*}
The respective Friedmann equations differ as well, i.e. for $k>0$
\begin{align}
 1 =\ &2\, \xi\, \Omega^2\, \left( 1 - \Omega_H^2 \right) + 3\, \xi\, \left( \sqrt{\frac{2}{3}}\, \Omega_H\, \Omega + \Omega_{\psi}\, \sqrt{1-\xi\, \Omega^2} \right)^2 \nonumber\\ 
 &+ (1-3\, \xi)\, \Omega_{\psi}^2\, \left(1-\xi\, \Omega^2\right) + \Omega_V^2\, \left(1-\xi\, \Omega^2\right) \, ,\label{eq:fried_pos}
 \end{align}
 while for $k\leq0$
 \begin{align}
 1 =\ &2\, \left(1-\xi\, \Omega^2\right)\, \left(1-\Omega_H^2\right) +  3\, \xi\, \left( \sqrt{\frac{2}{3}}\, \Omega_H\, \Omega + \Omega_{\psi}\, \sqrt{1-\xi\, \Omega^2} \right)^2\nonumber\\ 
 &+ (1-3\, \xi)\, \Omega_{\psi}^2\, \left(1-\xi\, \Omega^2\right) + \Omega_V^2\, \left(1-\xi\, \Omega^2\right)\, .\label{eq:fried_neg}
\end{align}
 
\subsection{General features of the system}

\paragraph{\textbf{Symmetries.}}
The dynamical system (\ref{eq:omega})-(\ref{eq:omegadv}) remains invariant under the simultaneous transformation 
\begin{align}\label{eq:symm}
    \{\Omega, \Omega_{H}, \Omega_{\psi}, \Omega_{V}, \Omega_{\partial V} \} \to \{-\Omega,\Omega_{H},-\Omega_{\psi},\Omega_{V},-\Omega_{\partial V} \} \,  .
\end{align}
This symmetry, physically, is equivalent to the invariance under the transformation $\psi \rightarrow -\psi$. Since $\Omega_V$ is not affected by this transformation~\eqref{eq:symm}, then it must hold that $V(\psi)= V(-\psi)>0$.

\paragraph{\textbf{Singularities.}}
In this system there are singular points arising from the decoupling of Raychaudhuri and Klein-Gordon equations, i.e. where the determinant of their Jacobian vanishes. These singular points, in terms of dimensionless variables, correspond to the vanishing of
\begin{equation} \label{eq:sing}
    \Omega=\pm \frac{1}{\sqrt{2\xi(1-3\xi)}}.
\end{equation}
By substituting the former relation into the Friedmann constraints and solving for $\Omega_{\psi}$ one gets
\begin{equation}
    \Omega_{\psi}=\frac{\sqrt{6\xi}\Omega_H+\sqrt{(\Omega_H^2\mp\Omega_V^2-1)6\xi\pm \Omega_V^2}}{\sqrt{1-6\xi}},
\end{equation}
where the upper/lower sign corresponds to negative/positive curvature. In the range $\xi \in (0,1/6)$, in both of these cases the coordinates $\left( \Omega, \Omega_{\psi} \right)$ of the singularity remain finite .  For $\xi>1/6$, $\Omega_\psi$ is complex.
In the case of a flat spacetime $\Omega_H=\pm 1$ we call these singularities $\mathcal{S}_{\pm}$ respectively. 

\paragraph{\textbf{Invariant subsets.}}
For the dynamical system.~\eqref{eq:omega}-\eqref{eq:omegadv}, one can identify some invariant subsets of the system. These invariant subsets are $\Omega_H=\pm1$ (flat spacetime) and $\Omega_V=0$ (free scalar field). 

\paragraph{\textbf{Critical points.}} Critical points and their physical interpretations of this system are summarized in the table~\ref{tab:CritPoints}.
\begin{table*}
  \caption{The critical elements of the system and their stability in the range $0\le\xi\le 1/6$.  
  }
  \begin{adjustbox}{width=.8\textwidth}
\small
\begin{tabular}{c | c c c c c | c c c c}\label{tab:CritPoints}
 & $\Omega_{\psi}$ &  $\Omega_H$ & $\Omega$ & $ \Omega_V$ & $\Omega_{\partial V}$ & Curvature & $q$ & $w_{e}$ & stability\\
\hline
 & & & & &  & & & & \\
  
  $\mathcal{A}_+$ & 0 & $1$ & 0 & $ 1$ & 0 & flat & -1 & -1 & sink\\
  $\mathcal{A}_-$ & 0 & $-1$ & 0 & $ 1$ & 0 & flat & -1 & -1 & source\\
  $\mathcal{B}_+$ & 0 & $1$ & $0<\Omega^2<\frac{1}{2\xi}$ & $ \sqrt{\frac{1-2\xi \Omega^2}{1-\xi \Omega^2}}$ & $-\frac{4\xi\Omega\sqrt{1-\xi\Omega^2}}{1-2\xi\Omega^2}$  & flat & -1 & -1 & sink\\
  $\mathcal{B}_-$ & 0 & $-1$ & $0<\Omega^2<\frac{1}{2\xi}$ & $\sqrt{\frac{1-2\xi \Omega^2}{1-\xi \Omega^2}}$ & $-\frac{4\xi\Omega\sqrt{1-\xi\Omega^2}}{1-2\xi\Omega^2}$  & flat & -1 & -1 & source\\
  $\mathcal{C}_{\pm}$ & 0 & $\pm 1$ & $\pm\frac{1}{\sqrt{2\xi}}$ & 0 & $\forall$ & flat & 1 & $\frac{1}{3}$ & saddle \\
  $\mathcal{D}_{\pm}$ & 0 & $\pm \frac{1}{\sqrt{2}}$ & $\forall$ & 0 & $\forall$  & negative & 0 & - 
     & saddle\\
\end{tabular}
\end{adjustbox}
\end{table*}
\section{Conclusions}\label{sec:concl}

This work introduces general frameworks to analyze dynamical systems of:
\begin{itemize}
    \item barotropic fluids with non-negative energy density and generic EoS,
    \item non-minimally coupled real scalar fields with generic potential in the absence of regular matter,
\end{itemize}
 both cases are treated in spatially curved FRW spacetimes without cosmological constant. In both cases we have employed a general $\Gamma$ function, which when specified reduces our general frameworks to specific models. We were able to identify critical elements and basic features of the systems for unknown $\Gamma$ functions.

\bibliographystyle{ragtime}
\bibliography{ker}

\end{document}